\documentclass[twoside,11pt]{article}

\usepackage{graphicx}

\usepackage{amssymb}
\usepackage{amsmath}
\usepackage{amsfonts}
\usepackage{wasysym}

\begin{document}           

\title{\Large 
Two-component perfect fluid in FRW universe
}
\author{V.E. Kuzmichev, V.V. Kuzmichev\\[0.5cm]
\itshape Bogolyubov Institute for Theoretical Physics,\\
\itshape National Academy of Sciences of Ukraine, Kiev, 03680 Ukraine}

\date{}

\maketitle

\begin{abstract}
We consider the cosmological model which allows to describe on equal footing the evolution of matter in the universe 
in the time interval from the inflation till the domination of dark energy. The matter has a form of a two-component 
perfect fluid imitated by homogeneous scalar fields between which there is energy exchange. 
Dark energy is represented by the cosmological  constant, which is supposed invariable during the whole evolution of the universe. 
The matter changes its equation of state with time, so that the era of radiation domination in the early universe smoothly passes into 
the era of a pressureless gas, which then passes into the late-time epoch, when the matter is represented 
by a gas of low-velocity cosmic strings. The inflationary phase is described as an analytic continuation 
of the energy density in the very early universe into the region of small negative values of the parameter 
which characterizes typical time of energy transfer from one matter component to another.
The Hubble expansion rate, energy density of the matter, energy density parameter, and deceleration parameter as functions of time
are found.
\end{abstract}


\section{Introduction}

The standard cosmological model \cite{PDG,Rio} describes the universe which contains relativistic matter (radiation), 
visible and dark matter, and dark energy in the form of the vacuum (cosmological constant) or quintessence 
(hypothetical cosmic fluid with the equation of state parameter which changes with time). The universe
evolves so that in the course of the expansion, one or another matter-energy component 
begins to dominate over the others.

In the early universe the dominating matter component is radiation with the energy density 
$\rho \sim R^{-4}$,  where $R$ is the cosmic scale factor.
Then, as the universe expands, the non-relativistic matter (dust) with the energy density $\rho \sim R^{-3}$
begins to dominate.
If one assumes that in the late-time universe there exists a complementary linear constraint between the cosmic scale factor and total amount of 
matter in the universe as a whole, the latter phase will pass into the stage in which the matter in the universe will look like a perfect gas of 
low-velocity cosmic strings with $\rho \sim R^{-2}$ \cite{Ku}.
In the radiation-dominated universe the number density of photons is $n_{\gamma} \sim R^{-3}$,
and the energy of every photon decreases, during the expansion of the universe,
as $m_{\gamma} \sim R^{-1}$ due to the cosmological redshift. 
As a result, the effective mass of the matter attributed to relativistic matter reduces as well,
$M_{eff} \sim m_{\gamma}  n_{\gamma}  R^{3} \sim R^{-1}$. Arguing in the same way, one finds that in 
the universe, in which the matter is represented by a dust, the effective mass is constant, $M_{eff} = const$, expressing the 
constancy of the sum of the masses of the bodies in the volume $\sim R^{3}$. 
In the low-velocity cosmic strings phase, the effective mass of the matter increases with the expansion of the universe,
$M_{eff} \sim R$. 
On the time interval from the reheating after the inflation till the domination of dark energy, 
the equation of state of the dominating matter changes from the equation for radiation with the pressure
$p = \frac{1}{3}\rho$ to the equation for the low-velocity cosmic strings
$p = - \frac{1}{3}\rho$ passing through the pressureless state $p = 0$.

It seems justified to consider the approach in which
the equation of state of the matter in the universe changes continuously in time, 
passing through the limiting cases being specified. Models, in which 
continuous change of state of the matter was assumed, were studied by a number of authors
\cite{Dym,Kra}.

In the present article we consider the model of the universe, in which the matter
has a form of a two-component perfect fluid, while dark energy is represented
by the cosmological constant, which is supposed invariable during the whole evolution of the universe.
Components of a fluid are described by spatially homogeneous scalar fields. The first component
has the equation of state which changes in time from the stiff Zel'dovich type equation to the vacuum type one. 
The second component is pressure-free matter.
During the course of the evolution of the universe, there is redistribution of energy between the components. 
As a result, the total matter changes its equation of state in time, so that
the era of radiation domination in the early universe smoothly passes into the era, when
the matter is represented by a pressureless gas. The latter changes into the epoch in which the matter is
described by a gas of low-velocity cosmic strings. The solution obtained for the 
era of radiation domination in the very early universe can be analytically continued into
the region of extremely small negative values of the parameter
which characterizes typical time of energy transfer from one matter component to another.
At the same time, the first matter component acquires the properties of the vacuum
with the corresponding equation of state and it can be identified with the inflaton field,
used in models of inflation, in this stage.
The contribution from the cosmological constant can be neglected in the early universe, 
while in the late-time epoch its role is vital to produce the accelerating expansion.

For the homogeneous, isotropic and spatially flat universe with the Friedmann-Robertson-Walker (FRW) metric,
in terms of a two-component model, we find the explicit expressions for the Hubble expansion rate, 
energy density of the matter, energy density parameter, and deceleration parameter as functions of time
for the whole time interval. The limiting cases corresponding to small, intermediate (when the matter in the universe is represented by a
dust), and large (the late-time epoch of low-velocity cosmic strings) values of time are considered for a comparison with
known solutions.

\section{Two-component perfect fluid}

Let us consider the FRW universe which is described by a set of equations
\begin{equation}\label{1}
H^{2} \equiv \left(\frac{\dot{R}}{R} \right)^{2} = \frac{8 \pi G}{3} \left(\rho + \rho_{\Lambda} \right),
\end{equation}
\begin{equation}\label{2}
\dot{\rho} + 3 H (\rho + p) = 0,\qquad \dot{\rho}_{\Lambda} = 0,
\end{equation}
\begin{equation}\label{3}
p = w(t)\, \rho,\qquad p_{\Lambda} = - \rho_{\Lambda},
\end{equation}
where $R(t)$ is the cosmic scale factor, $\rho(t)$ is the energy density of the matter which has a form of the 
homogeneous perfect fluid, $p(t)$ is its pressure, $w(t)$ is the equation of state parameter,
$\rho_{\Lambda}$ is the energy density of the vacuum which is connected with the cosmological constant 
$\Lambda$ in the usual way as $\rho_{\Lambda} \equiv \frac{\Lambda}{8 \pi G}$,
$p_{\Lambda}$ is its pressure,
$G$ is the Newtonian gravitational constant, an overdot denotes $d/dt$, $t$ is the proper time 
(units $c = 1$ are used). 

We assume that the matter consists of two components,
\begin{equation}\label{4}
\rho = \rho_{b} + \rho_{d},\qquad p = p_{b} + p_{d}.
\end{equation}
It is supposed that a redistribution of energy between these components takes place 
in the course of the evolution of the universe. The energy conservation equation for the matter (\ref{2}) rewritten for components
takes the form
\begin{equation}\label{5}
\dot{\rho}_{b} + 3 H (\rho_{b} + p_{b}) = Q,\quad \dot{\rho}_{d} + 3 H (\rho_{d} + p_{d}) = -Q,
\end{equation}
where $Q$ describes the interaction between the components.

The components of the perfect fluid are imitated by scalar fields $\phi_{b} (t)$ and $\phi_{d} (t)$
with potentials $V_{b}(\phi_{b})$ and $V_{d}(\phi_{d})$,
\begin{equation}\label{6}
\rho_{i} = \frac{1}{2} \dot{\phi_{i}}^{2} + V_{i},\qquad p_{i} = \frac{1}{2} \dot{\phi_{i}}^{2} - V_{i},
\qquad i = \{b,d\}.
\end{equation}
The models of such a type which include a coupling between the matter components
were considered in the literature, in particular, within the context of inflation and reheating
and the coincidence problem of dark energy and matter in the accelerating universe 
(see, e.g., Refs. \cite{Bil,Ame,Zim} and references therein). 
The form of the interaction term $Q$ may be derived from different physical arguments or obtained as a solution 
of some dynamical equation, which describes the required properties of the matter fields $\phi_{i}$.
Since, according to Eqs. (\ref{1})-(\ref{3}), dark energy is modeled by the cosmological constant,
only the evolution of the matter as a two-component perfect fluid is considered below.

Let us assume that the field $\phi_{d}$ forms the pressure-free matter component (dust),
\begin{equation}\label{7}
\frac{1}{2} \dot{\phi_{d}}^{2} = V_{d},\qquad \rho_{d} = 2V_{d},\qquad p_{d} = 0.
\end{equation}
The potential $V_{d}$ is taken equal to the density $\rho_{b}$, $V_{d} = \rho_{b}$. Then,
$\rho_{d} = 2 \rho_{b}$, and 
the total energy density, the pressure, and the parameter $w$ are
\begin{equation}\label{8}
\rho = 3 \rho_{b}, \qquad p = p_{b},\qquad w = \frac{p_{b}}{3 \rho_{b}}.
\end{equation}
In this case, $Q = 2 H p_{b}$ and the set of equations (\ref{5}) reduces to one equation
\begin{equation}\label{9}
\dot{\rho}_{b} + 3 H \left(\rho_{b} + \frac{1}{3} p_{b}\right) = 0.
\end{equation}
The following condition is imposed on the second field $\phi_{b}$,
\begin{equation}\label{10}
\frac{1}{2} \dot{\phi_{b}}^{2} = V_{b}\, \mbox{e}^{- 2 \left(t - t_{0}\right) / \tau},
\end{equation}
where $t_{0}$ is the instant of time close to which the field $\phi_{b}$ 
reproduces the pressure-free matter, $\tau$ is the parameter which characterizes
some typical time of energy transfer from one component to another and the value
$1/\tau$  determines the mean rate of change of the equation of state parameter $w(t)$ of Eq. (\ref{8}).
 
The field $\phi_{b}$, which satisfies the condition (\ref{10}), describes the matter component
which evolves in time from the stiff Zel'dovich matter,
\begin{equation}\label{11}
p_{b} \approx \rho_{b} \quad \mbox{at} \ t \approx 0, \quad \mbox{for}\ \frac{t_{0}}{\tau} > 2,
\end{equation}
through the dust,
\begin{equation}\label{12}
p_{b} \approx 0 \quad \mbox{at} \ t \approx t_{0},
\end{equation}
to the matter with vacuum-type equation of state,
\begin{equation}\label{13}
p_{b} \approx - \rho_{b} \quad \mbox{at} \ t \gg t_{0}.
\end{equation}
Substituting Eq. (\ref{10}) into (\ref{6}) and using the expression for $w$ (\ref{8}), we find 
\begin{equation}\label{15}
w(t) = - \frac{1}{3} \tanh \left(\frac{t - t_{0}}{\tau}\right).
\end{equation}
Thus, in the model under consideration, the equation of state parameter for the two-component matter $w$ 
changes from $\frac{1}{3}$, when Eq. (\ref{11}) holds, passing through the point $w = 0$ for the pressure-free matter component 
(\ref{12}), to $-\frac{1}{3}$ in the case (\ref{13}), taking all intermediate values. 

Using the mechanical analogy, the function (\ref{15}) can be considered as the antikink solution
of the equation
\begin{equation}\label{14}
\frac{1}{2} \dot{w}^{2} + [- U(w)] = 0, \quad U = \frac{9}{2 \tau^{2}} \left(w^{2} - \frac{1}{9}\right)^{2},
\end{equation}
which describes the motion of the analogue particle with zero energy in the potential $[- U(w)]$ (cf., e.g., Ref. \cite{Raj}).
This potential has two maxima at the points $w = \pm \frac{1}{3}$ and a local minimum at $w = 0$. 
The analogue particle moves along the `trajectory' (\ref{15}) from the value $w = \frac{1}{3}$ in the distant past ($t = - \infty$)
to the value $w = - \frac{1}{3}$ reached at $t = \infty$. At the moment $t = t_{0}$, the particle passes through the minimum
of the potential at $w = 0$. Leaving the point $w = \frac{1}{3}$, the analogue particle can only approach the point $w = - \frac{1}{3}$
at $t \rightarrow \infty$, where its velocity and acceleration vanish. It cannot return back to $w = \frac{1}{3}$. This analogy with the motion of 
the particle allows to understand a unidirectional evolution of matter in the universe governed by Eq. (\ref{1}) 
from the radiation-dominated era through dust domination
to the hypothetical late-time epoch, in which the matter is described by a gas of low-velocity cosmic strings. 
We note that Eq. (\ref{14}) has another solution in the form of the kink which is equal to the function (\ref{15}) with an inverse sign.
This case would correspond to the model in which a gas of low-velocity cosmic strings at $t = - \infty$ transforms into radiation at $t = \infty$,
but we do not consider it in the present paper.

Let us find, how the cosmological parameters, namely the Hubble expansion rate
and the total energy density, depend on time. 
From Eqs. (\ref{1}), (\ref{9}) and (\ref{15}), it follows the non-linear equation for the Hubble expansion rate,
\begin{equation}\label{16}
\dot{H} + \frac{1}{2} \left\{3 - \tanh \left[\frac{1}{\tau}\left(t - t_{0}\right)\right]\right\} 
\left(H^{2} - \frac{\Lambda}{3} \right) = 0.
\end{equation}
The general solution of this equation is
\begin{equation}\label{17}
H(t) = \sqrt{\frac{\Lambda}{3}} \coth Z\left(t\right),
\end{equation}
where we denote
\begin{equation}\label{18}
Z \left(t\right) \equiv \frac{1}{2} \sqrt{\frac{\Lambda}{3}}
\left[C t_{0} + 3 t - \tau \ln \cosh \left( \frac{1}{\tau}\left(t - t_{0}\right)
\right)\right],
\end{equation}
$C$ is the constant of integration.

Substituting the solution (\ref{17}) into Eq. (\ref{1}), we find the energy density of the matter as
a function of time
\begin{equation}\label{19}
\rho(t) = \rho_{\Lambda} \sinh^{-2} Z \left(t\right).
\end{equation}
The energy density parameter of the matter, $\Omega_{M} = \rho / (\rho + \rho_{\Lambda})$, is
\begin{equation}\label{20}
\Omega_{M} (t) = \cosh^{-2} Z \left(t\right).
\end{equation}
The deceleration parameter, $q = -1 - \dot{H}/H^{2}$, is equal to
\begin{equation}\label{21}
q (t) = -1 + \frac{3 - \tanh \left[\left(t - t_{0}\right)/\tau\right]}{2 \cosh^{2} Z \left(t\right)}.
\end{equation}

The scalar fields $\phi_{b}$ and $\phi_{d}$ themselves can be restored by integrating the expressions
\begin{equation}\label{22}
\phi_{b} = \sqrt{\frac{2}{3}} \int\!\!dt \sqrt{\frac{\rho(t)}{\mbox{e}^{2 \left(t - t_{0}\right) / \tau} + 1}},\quad
\phi_{d} = \sqrt{\frac{2}{3}} \int\!\!dt \sqrt{\rho(t)}.
\end{equation}
Their potentials are
\begin{equation}\label{23}
V_{b}(t) = \frac{1}{3} \frac{\rho(t)}{\mbox{e}^{-2 \left(t - t_{0}\right) / \tau} + 1},\quad
V_{d}(t) = \frac{1}{3} \rho(t).
\end{equation}

\section{Limiting cases}

Let us consider the different limiting cases, when the matter takes the forms of radiation ($t \ll t_{0}$),
dust ($t \sim t_{0}$), and a perfect gas of low-velocity cosmic strings ($t \gg t_{0}$). 

1. \textit{The late-time epoch of low-velocity cosmic strings}.

The Hubble expansion rate, the energy density, and the deceleration parameter
in this case do not depend on the choice of 
the constant of integration $C$ and the parameters $t_{0}$ and $\tau$,
\begin{eqnarray}\label{24}
& & H (t) = \sqrt{\frac{\Lambda}{3}} \coth \left(\sqrt{\frac{\Lambda}{3}} t \right),
\quad \rho(t) = \rho_{\Lambda} \sinh^{-2} \left(\sqrt{\frac{\Lambda}{3}} t \right),
\nonumber \\
& & \Omega_{M} (t) = \cosh^{-2} \left(\sqrt{\frac{\Lambda}{3}} t \right),
\quad q(t) = -\tanh^{2} \left(\sqrt{\frac{\Lambda}{3}} t \right)
\quad \mbox{at} \ t \gg t_{0}.
\end{eqnarray}
If the cosmological constant is equal to zero, $\Lambda = 0$, then
\begin{equation}\label{25}
H (t) = \frac{1}{t},\quad \rho(t) = \frac{3}{8 \pi G t^{2}}, \quad q = 0.
\end{equation}
The equations (\ref{24}) reproduce the result of Ref. \cite{Ku}. The equation for
$\rho(t)$ (\ref{25}) has a form of Whitrow-Randall's relation \cite{Whi}.

2. \textit{The epoch, when the matter is in the form of dust}.

In this epoch the parameters are
\begin{eqnarray}\label{26}
& & H (t) = \sqrt{\frac{\Lambda}{3}} \coth \left(\frac{1}{2} \sqrt{\frac{\Lambda}{3}} 
\left(3 t + C t_{0} \right) \right),
\quad \rho(t) = \rho_{\Lambda} \sinh^{-2} \left(\frac{1}{2} \sqrt{\frac{\Lambda}{3}} 
\left(3 t + C t_{0} \right) \right),
\nonumber \\
& & \Omega_{M} (t) = \cosh^{-2} \left(\frac{1}{2} \sqrt{\frac{\Lambda}{3}} 
\left(3 t + C t_{0} \right) \right),
\quad q(t) = -1 + \frac{3}{2}\, \Omega_{M} (t)
\quad \mbox{at} \ t \sim t_{0}.
\end{eqnarray}
Choosing the constant of integration $C = 0$, one restores the expressions obtained in Ref. \cite{Ku,Gr}.
In the case $\Lambda = 0$ and $C = 0$, we have the following well-known result \cite{LL},
\begin{equation}\label{27}
H (t) = \frac{2}{3 t},\quad \rho(t) = \frac{1}{6 \pi G t^{2}},\quad q = \frac{1}{2}.
\end{equation}

3. \textit{The epoch of radiation}.

We have
\begin{eqnarray}\label{28}
& & H (t) = \sqrt{\frac{\Lambda}{3}} \coth \left(\frac{1}{2} \sqrt{\frac{\Lambda}{3}} 
\left[4 t + \left(C - 1\right) t_{0}  + \tau \ln 2 \right] \right),
\nonumber \\
& & \rho(t) = \rho_{\Lambda} \sinh^{-2} \left(\frac{1}{2} \sqrt{\frac{\Lambda}{3}} 
\left[4 t + \left(C - 1\right) t_{0}  + \tau \ln 2 \right] \right),
\nonumber \\
& & \Omega_{M} (t) = \cosh^{-2} \left(\frac{1}{2} \sqrt{\frac{\Lambda}{3}} 
\left[4 t + \left(C - 1\right) t_{0}  + \tau \ln 2 \right] \right),
\nonumber \\
& & q(t) = -1 + 2\, \Omega_{M} (t)
\quad \mbox{at} \ t \ll t_{0}
\end{eqnarray}
Choosing $C = 1 - \frac{\tau}{t_{0}} \ln2$, in the case of $\Lambda = 0$, we obtain the known result 
\cite{LL}
\begin{equation}\label{29}
H(t) = \frac{1}{2t}, \quad \rho (t) = \frac{3}{32 \pi G t^{{2}}}, \quad q = 1.
\end{equation}
In the special case, when $t \ll \frac{1}{4} |\left(C - 1\right) t_{0}  + \tau \ln 2 |$ (it includes the very early universe with
$t \gtrsim M_{p}^{-1}$, where $M_{p} = G^{-1/2}$ ($\hbar = c = 1$) is the Planck mass),
it follows from Eq. (\ref{28}) that the energy density $\rho$
remains almost constant and equals to
\begin{equation}\label{30}
\rho_{e} = \rho_{\Lambda} \sinh^{-2} \left(\frac{1}{2} \sqrt{\frac{\Lambda}{3}} 
\left[\left(C - 1\right) t_{0}  + \tau \ln 2 \right] \right).
\end{equation}
Passing to the limit $\Lambda = 0$, we obtain
\begin{equation}\label{31}
\rho_{e} = \frac{3}{2 \pi G t_{0}^{2}} \frac{1}{\left(C - 1 + \frac{\tau}{t_{0}} \ln 2\right)^{2}}.
\end{equation}

4. \textit{The era of inflation}.

Concerning the evolution of the matter, the era of inflation demands special examination.
In the approach under consideration,
we can perform an analytic continuation in the expression (\ref{31}) into the region of small negative values of the parameter 
$\tau$, $\tau < 0$, $|\tau| \ll t_{0}$.
Then, it follows from Eq. (\ref{10}) that $\dot{\phi_{b}} \approx 0$ and the corresponding matter component 
has the vacuum-type equation of state, $p_{b} \approx - \rho_{b} \approx - V_{b}$. It can be identified with the scalar 
field (inflaton) used in models of inflation to describe the evolution of the very early universe \cite{Rio,Ly}.
Really, in this epoch, the energy density is $\rho = 3 \rho_{b} \approx 3 V_{b}$, where
\begin{equation}\label{32}
V_{b} = \frac{1}{2 \pi G t_{0}^{2}} \frac{1}{\left(C - 1  - \frac{|\tau|}{t_{0}} \ln 2\right)^{2}}
\end{equation}
plays the role of the inflaton potential.
Assuming that the inflationary era is smoothly connected to the radiation-dominated phase,
from Eqs. (\ref{31}) and (\ref{32}) it follows that the constant $C$ must be close to unity, so that the potential
(\ref{32}) is very large. 
Setting $C = 1$ and $|\tau| \approx M_{p}^{-1}$, we can rewrite Eq. (\ref{32}) in a simple form $V_{b} \approx \frac{1}{3}\, M_{p}^{4}$.
Substituting $\rho \approx M_{p}^{4} \gg \rho_{\Lambda}$ into Eq. (\ref{1}), we find the solution
$R(t) \sim \exp \left(\sqrt{\frac{8 \pi}{3}} M_{p} t \right)$, which describes inflation.

In the model under consideration, the inflationary era is driven by one of the components
of the two-component matter, while the cosmological constant remains invariable at all stages of the evolution of the universe.
In such an approach, the contribution from the cosmological constant can be neglected in the early universe, 
while in the late-time epoch of accelerating expansion its role becomes crucial, according to Eqs. (\ref{24}) and (\ref{26}).

\section{Discussion}

\begin{figure*}
\includegraphics[width=12cm]{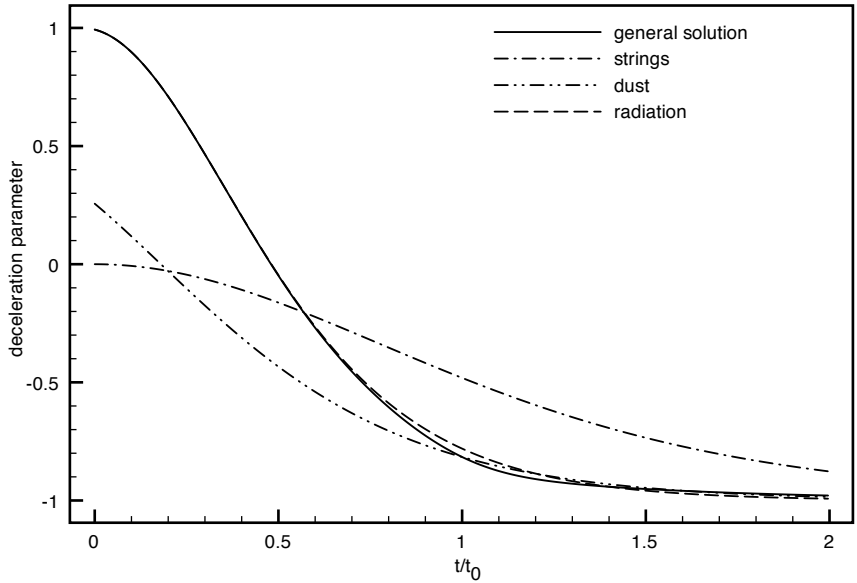}
\caption{The decelerating parameters (\ref{21}), (\ref{24}), (\ref{26}), and (\ref{28}) versus time. 
The value of
$\sqrt{\frac{\Lambda}{3}} t_{0} = 0.855$ following from the WMAP7 data \cite{Lar} is used.
For definiteness, the constants $C = 1$ and $\tau = 0.2 t_{0}$ are chosen.}
\end{figure*} 

In Fig.~1, it is shown the time dependence of the decelerating parameter calculated in accordance with the general solution 
(\ref{21}) and in cases when the relations for the epochs of low-velocity cosmic strings (\ref{24}), dust
(\ref{26}), and radiation (\ref{28}) are extrapolated to the whole time interval. 
The free parameters $\Lambda$ and $t_{0}$ (as the age of the universe) were taken from the WMAP7 data \cite{Lar}.
All curves go to $-1$ for infinite $t/t_{0}$.
The deceleration of the universe turns into acceleration at the instant of time $t \approx 0.48 t_{0}$. 
If $t_{0}$ is identified with the age of the universe,
the corresponding cosmic redshift is $z \approx 0.28$. It can be seen from Fig.~1 that,
near the instant of time $t = t_{0}$, 
the curves for a dust and radiation are close to the curve corresponding to the general solution (\ref{21}),
whereas the curve for strings is far away from it. This can be interpreted as evidence that in the region
$t \sim t_{0}$, effectively, the dominating matter component
is a dust with a small addition of the radiation, while a gas of cosmic strings makes a negligible influence on the dynamics 
of the universe here.

For a two-component perfect fluid with vanishing cosmological constant, $\Lambda = 0$,
the solution of Eq. (\ref{16}) has a form
\begin{equation}\label{33}
H \left(t\right) = 2
\left[C t_{0} + 3 t - \tau \ln \cosh \left( \frac{1}{\tau}\left(t - t_{0}\right)
\right)\right]^{-1}.
\end{equation}
In this case the deceleration parameter does not depend on the constant of integration $C$,
\begin{equation}\label{34}
q (t) = \frac{1}{2} \left\{1 - \tanh\left[\left(t - t_{0}\right)/\tau\right]\right\}.
\end{equation}
This function is plotted in Fig~2. The deceleration parameter changes from $1$ to $0$ as $t/t_{0}$ changes from $0$
to infinity. 
Between the radiation-dominated phase ($t/t_{0} \lesssim 0.5$) and the era of cosmic string gas domination ($t/t_{0} \gtrsim 1.5$),
there is the transitional domain, where the phase of the radiation domination smoothly passes into the era of dust domination 
($t/t_{0} \sim 1$), which, in turn, then passes into the late-time epoch of cosmic string gas domination. Due to the gravitational 
attraction of the matter, the universe will be decelerating on the time interval from $t/t_{0} = 0$ to $t/t_{0} \sim 1.5$.
From the instant of time $t/t_{0} \sim 1.5$ to infinity, the effective gravitational attraction of the perfect gas of low-velocity cosmic strings
will be practically compensated by its negative pressure acting repulsively, and such a universe will expand with the constant velocity.
In the region of the large values of $t/t_{0}$, the cosmic scale factor depends on time as $R \sim t$.
As it is claimed, e.g. in Ref. \cite{Gum}, power-law cosmology, $R \sim t^{\alpha}$ with $\alpha \approx 1$, 
is consistent with the recent Wilkinson Microwave Anisotropy Probe (WMAP7), Baryon Acoustic Oscillations (BAO) and Hubble data,
therefore the consideration of models, in which the cosmic scale factor evolves linearly with time, while
the cosmological constant is negligible, may be of practical interest.

\begin{figure*}
\includegraphics[width=12cm]{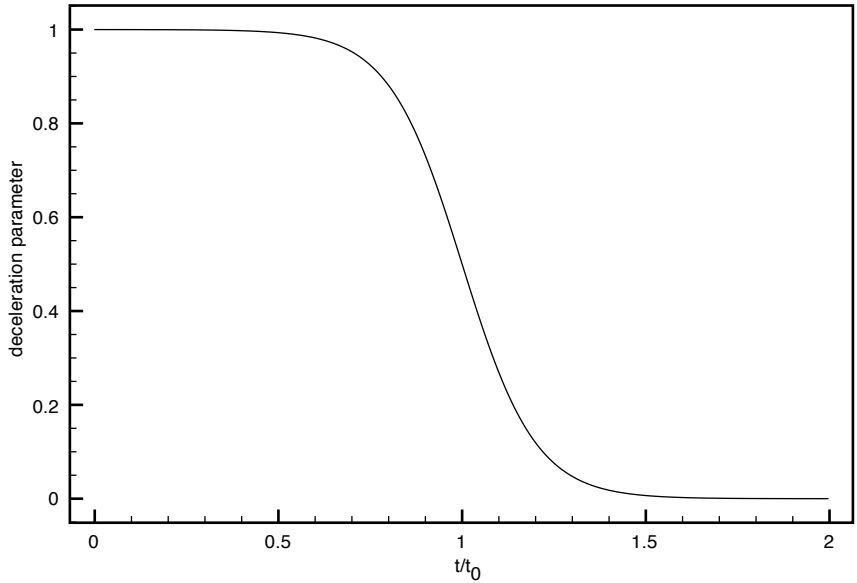}
\caption{The decelerating parameter (\ref{34}) versus time. 
It is taken $\tau = 0.2 t_{0}$.}
\end{figure*} 

Comparing Fig.~1 with Fig.~2, one finds that the general behavior of the deceleration parameter with time remains the same,
however, the action of the cosmological constant smoothes the curve and also produce a new physical effect 
which results in a change of the sign of the deceleration parameter, so that the deceleration phase in the history of the universe
passes into subsequent period of accelerating expansion.

\section{Conclusion}

In the present paper, we propose a single approach to the description of the evolution of the matter in the universe with
non-zero cosmological constant in the whole time interval from the inflation till the domination of dark energy.
The new element here is the inclusion of the hypothetical late-time epoch, when the matter is represented 
by a gas of low-velocity cosmic strings, in the general scheme. In some sense, these strings may be considered
as emergent ones, since they correspond to the collective motion of the matter. We draw attention to the fact that, 
if the final stage of the evolution of the matter in the universe is not a pressure-free matter, but a gas of
low-velocity cosmic strings, then there arises an additional `symmetry' in the equation of state of the matter.
Namely, the equation of state parameter $w$ (\ref{15}) changes from $\frac{1}{3}$, passing through the point $w = 0$, 
to $-\frac{1}{3}$, taking all intermediate values. For such a model, we find the solution (\ref{17}) of the equation for the 
Hubble expansion rate (\ref{16}) in analytical form as a function of time. The Hubble expansion rate (\ref{17}), 
the energy density of the matter (\ref{19}), and the deceleration parameter (\ref{21}) as functions of time 
reproduce all known cosmological solutions (\ref{25}), (\ref{27}), and (\ref{29}) for the different limiting cases
obtained by a number of authors \cite{Ku,Whi,Gr,LL} for vanishing cosmological constant.

\end{document}